\documentclass[a4paper,showpacs,11pt]{article}
\usepackage{amsfonts}
\usepackage{amsmath,amssymb}
\usepackage{graphicx}
\title{Applications of the potential algebras of the two-dimensional Dirac-like operators}
\author{\textsf{V\'{\i}t Jakubsk\'y}
 \\
{\small \textit{Nuclear Physics Institute, \v Re\v z near Prague, 25068, Czech Republic}}\\
 \\
{\small \textit{E-mail: v.jakubsky@gmail.com
} }\\
\date{}
}
\begin{document}

\maketitle

\begin{abstract}
Potential algebras can be used effectively in the analysis of the quantum systems.
In the article, we focus on the systems described by a separable, $2\times 2$ matrix Hamiltonian of the first order in derivatives.
We find integrals of motion of the Hamiltonian that close centrally extended
$so(3)$, $so(2,1)$ or oscillator algebra. The algebraic framework is used
in construction of physically interesting solvable models described by the $(2+1)$ dimensional Dirac equation.
It is applied in description of open-cage fullerenes where the energies
and wave functions of low-energy charge-carriers are
computed. The potential algebras are also used in construction of shape-invariant, 
one-dimensional Dirac operators. We show that shape-invariance of the 
first-order operators is associated with the $N=4$  nonlinear
supersymmetry which is represented by both local and nonlocal supercharges. The relation to the shape-invariant
non-relativistic systems is discussed as well.   
\end{abstract}
 
\section{Introduction}
Exactly solvable models play an exceptional role in physics. They are simple 
enough to be solvable, yet sufficiently complicated to grab the essence of 
physical reality.  In quantum mechanics, exact solvability usually means 
that the eigenfunctions and eigenvalues of the Hamiltonian can be found 
explicitly. They can be computed analytically by solving  directly  
corresponding differential equations. 

Alternatively, they can be obtained 
in algebraic manner with the use of the integrals of motion; when 
the symmetries of the quantum system form Lie algebra, the spectrum and  
eigenstates of the energy operator  can be deduced from the analysis of 
the admissible unitary representations \cite{sga1}, \cite{sga2}. 
The symmetry operators preserve 
domain of the Hamiltonian by transforming one physical state into another. 
When they are time-dependent, they do not commute 
with the energy operator and, hence, they do 
not  preserve energy. The associated algebra of 
the integrals of motion is called dynamical or  spectrum generating; see 
\cite{Dothan}, \cite{sga3}. When the integrals of motion 
commute with the Hamiltonian, they rather reflect 
spectral degeneracy of the system. The algebraic structure, where 
the Hamiltonian plays the role of central element, is denoted as potential algebra.   
In the current article, 
we will focus on the analysis and applications of the potential algebras of the 
relativistic quantum systems described by the $(2+1)$ dimensional Dirac 
equation.

The low-dimensional Dirac equation appears in surprising 
variety of physical settings that are studied both in quantum field theory 
and in the condensed matter physics. Let us mention the $(1+1)$ 
dimensional variant of the famous Nambu-Jona-Lasinio (chiral Gross-Neveu) 
model \cite{Dunne}, \cite{Feinberg}, \cite{Thies}, analysis of 
fractionally charged solitons \cite{Jackiw}, \cite{Jackiw2} as well as 
description of inhomogeneous superconductors \cite{inhomogeneous}, carbon 
nanotubes \cite{KaneMele}, \cite{twisting}, or linear molecules like 
polyacetylene \cite{Takayama}.  The $(2+1)$ dimensional massless Dirac 
Hamiltonian appears in 
the low-energy approximation of dynamics of charge carriers in graphene 
and in related carbon nanostructures \cite{Wallace}, \cite{Semenoff}, 
\cite{Osipov}, \cite{VozmedianoGauge}.  

The work is organized as follows: in the next section, the potential 
algebras are constructed for a separable Dirac-like operator with the use 
of a specific ansatz for both the ladder operators and for the structure of 
the algebra. Relevant aspects of the representations based on the 
lowest/highest weight vectors are discussed in more detail, as well as the application  
of the algebraic structure in description of Dirac fermions in the 
curved space. 

The results are used directly in the third section, where solvable 
models describing low-energy charge carriers on 
the surface of open-cage fullerenes are analyzed. Two configurations are considered, 
with one and two holes in the surface of the crystal.

In the fourth section, we apply the potential algebras in construction of 
one-dimensional shape-invariant systems. First, we focus on the shape-invariance 
of one-dimensional Dirac Hamiltonians and show that it is 
associated with $N=4$ nonlinear supersymmetry, generated by both local and non-local supercharges. We also discuss how is the relation of this framework with the known, non-relativistic shape-invariant Hamiltonians. In this context, we consider superalgebraic structure associated with the nonrelativistic shape-invariant systems whose supercharges are shape-invariant themselves. We make a comment on the 
nonrelativistic systems with position dependent mass and present an illustrative example of such system which possesses shape-invariance. 

In the last section, we discuss briefly  
two specific two-dimensional systems; we make few comments on Dirac 
oscillator and its dynamical symmetries. Additionally, we show that the 
$so(2,1)$ potential algebra can be used for analysis of $(1+1)$-
dimensional quantum system whose metric can be identified with the 
(restricted) BTZ black hole metric. We conclude with short discussion of the results and
outlook to possible future research.

\section{Potential algebras of two-dimensional Dirac operator}
Let us consider the operator  $h_D$ which is given in terms of the Pauli matrices  $\sigma_a$
and real functions $g_2$ and $g_3$,
\begin{equation}\label{hD}\nonumber
 h_D=i\sigma_1\partial_{x_1}+\sigma_2\left(-ig_2\partial_{x_2}+g_3\right)+M\sigma_3, \quad g_a=g_a(x_1),\quad a=2,3.
\end{equation}
$M$ is a real constant.
We leave the domain of $h_D$ as well as the range of 
the real variables $x_1$ and $x_2$ unspecified at the moment. The function $g_2$ should not be identically zero. If not stated 
otherwise, the coordinates $x_1$ and $x_2$ are considered to be space-like. 
The operator (\ref{hD}) can play the role of the Hamiltonian of the $(2+1)$-dimensional 
Dirac equation that governs dynamics of many physically interesting models mentioned in the previous section.  The vector potential $g_3=g_3(x_1)$ with real coupling constant $c_3$ is associated with an external magnetic 
field.  The operator is manifestly 
Hermitian with respect to the standard scalar product. 

The system represented by (\ref{hD}) has integral of motion $J_3$,
\begin{equation}\label{J3}
 J_3=-i\partial_{x_2},\quad [h_D,J_3]=0.
\end{equation}
We are interested in the settings where two additional integrals of motion 
$J_{\pm}$ exist, such that 
\begin{equation}\label{algebraansatz}
[J_3,J_{\pm}]= \pm J_{\pm},\quad [J_-,J_+]= 2c_1 J_3+c_2,\quad  [J_{\pm},h_D]=0, 
\end{equation}  
where $c_1$ and $c_2$ are real numbers the value of which will be specified later in the text. Let us notice that for $c_1\neq 0$ the non-vanishing $c_2$ just implies a constant shift in definition of the angular operator. 

Let us make the ansatz in the form $J_{+}=e^{ix_2}\left(\sum_{k=1}^2(A_{k,1}+
\right.$ $\left.\sigma_3A_{k,2})\partial_{x_k}\right.$ $\left.+C_1+\sigma_3 C_2\right)$ 
where the coefficients $A_{k,a}$ and $C_{a}$ ($a,k=1,2$) depend on 
$x_1$ only. The operator $J_-$ is defined as $J_-=J_{+}^{\dagger}$. 
Inserting the ansatz into (\ref{algebraansatz}) and comparing the 
coefficients of corresponding derivatives, we find that the  coefficients 
of $J_{\pm}$ can be expressed in terms of $g_2$ and 
$g_3$ as
\begin{equation}\label{Jpm}
 J_{\pm}=ie^{\pm ix_2}\left(\partial_{x_1}+\frac{g_2'(\pm J_3+\frac{1}{2})}{g_2}\pm\frac{g_3'}{g_2}\mp\frac{g_2}{2}\sigma_3\right).
\end{equation}
It suggests that the function $g_2$ should be node-less to avoid 
singularities in definition of these operators.
Additionally, the functions $g_2$ and $g_3$ have to solve the 
following set of differential equations,
\begin{eqnarray}
\label{system1}
g_2^4+c_1g_2^2&=&g_2'^2\,,\\
g_2^4+g_2'^2&=&g_2g_2''\,,\label{system15}\\
g_2^3g_3+\frac{c_2}{2}g_2^2&=&g_3'g_2'\,.\label{system2}
\end{eqnarray}
Quick inspection suggests that the system can be reduced to two equations as long as 
$g_3=g_2$. However, we will omit this solution; looking at (\ref{hD}),
it would just shift the operator $J_3$ by a constant factor. 
The equation (\ref{system15}) eliminates constant (nonzero) solution for $g_2$. 

We can find the following nontrivial solutions of the system, 
\begin{equation}\label{g2}\nonumber
 g_2=\left\{\begin{array}{l}\sinh^{-1} x_1,\\\\
\cos^{-1} x_1,\\ \\ x_1^{-1},\end{array}\right.\quad\mbox{and}\quad 
g_3=\left\{\begin{array}{l}c_3\coth x_1,
\quad \mbox{for}\quad c_1=1,\ c_2=0,\\\\
c_3\tan x_1,\quad \mbox{for}\quad c_1=-1,\ c_2=0, \\ \\ c_3x_1,\quad \mbox{for}\quad c_1=0,\ c_2=-4c_3,\end{array}\right.
\end{equation}
where $c_3$ is a real coupling constant. To avoid singularities, the variable $x_1$ can acquire only nonzero values for $c_1\in\{1,0\}$ and $x_1\in\left(-\frac{\pi}{2},\frac{\pi}{2}\right)$ for $c_1=-1$.
The equations (\ref{system1})-(\ref{system2}) are invariant with respect to reflections $x_1\rightarrow -x_1$ as well as they possess translational invariance. It means that the functions (\ref{g2}) solve the system even after a constant shift and reflection of the coordinate $x_1$.
It is possible to write the potential term $g_3$ as the function of $g_2$ and $c_1$,
\begin{equation}
  g_3^2=c_3^2(g_{2}^2+c_1),
\quad \mbox{for}\quad c_1\neq0,\qquad g_3^2=c_3^2g_{2}^{-2},\quad \mbox{for}\quad c_1=0.
\end{equation} 

The constants $c_1$ and $c_3$ fix the structure of the algebra (\ref{algebraansatz}).
For $c_1\neq0$, it can be identified 
either as $so(3)$ ($c_1=-1$) or $so(2,1)$ ($c_1=1$), while the oscillator 
algebra is restored for $c_1=0$. $h_D$ commutes with both 
$J_3$ and $J_{\pm}$, and, hence, it is the central element of the following algebra
\begin{equation}\label{algebra1}
[J_3,J_{\pm}] =\pm J_{\pm},\quad  [J_{-},J_{+}]=\left\{\begin{array}{rl}2c_1J_3& \quad \mbox{for}\quad c_1\neq 0,\\-4c_3 &\quad \mbox{for}\quad c_1=0,\end{array}\right.
\end{equation}
\begin{equation}\label{algebracentralelement}
 [h_D,J_{\pm}]=[h_D,J_3]=0.
\end{equation}
It is worth noticing that nonrelativistic systems with similar algebraic 
background were classified as superintegrable in \cite{Nikitinsuperintegrable}.
The representation space of 
(\ref{algebra1})-(\ref{algebracentralelement}) is spanned by the 
eigenvectors of $h_D$. It can be decomposed into subspaces where $h_D$ 
acquires constant value and where the algebra 
(\ref{algebra1})-(\ref{algebracentralelement}) is irreducible. These 
subspaces can have either finite or infinite dimension, dependently on 
the type of the algebra.

The square of (\ref{hD}) can be written as a second-order polynomial 
in the generators of (\ref{algebra1}). When $c_1\neq 0$, it 
coincides  
up to an additive constant with the standard Casimir operator $\mathcal{C}=J_+J_--
c_1J_3(J_3-1)=J_-J_+-c_1J_3(J_3+1)$ of the algebra $so(3)$ or $so(2,1)$. When $c_1=0$, the algebra is not semi-simple 
(Killing form is degenerate) and the standard (quadratic) Casimir operator 
does not exist. There holds 
\begin{equation}\label{hDsquare}
 h_D^2-M^2=\left\{\begin{array}{ll}J_+J_--c_1\left(\left(J_3-\frac{1}{2}\right)^2-c_3^2\right)=J_-J_+-c_1\left(\left(J_3+\frac{1}{2}\right)^2-c_3^2\right)& \quad \mbox{for}\quad c_1\neq 0,\\ J_+J_-+4c_3\left(J_3-\frac{1}{2}\right)=J_-J_++4c_3\left(J_3+\frac{1}{2}\right) &\quad \mbox{for}\quad c_1=0.\end{array}\right.
\end{equation}

For any of the solutions (\ref{g2}), the operator $h_D$ commutes with the linear operator $\mathcal{P}=\sigma_3\,R_{x_1}$ where $R_{x_1}$ is defined as $R_{x_1}x_1R_{x_1}=-x_1$ and $R_{x_1}\,x_2\,R_{x_1}=x_2$. The ladder operators $J_{\pm}$ together with $J_3$ satisfy the following relations, 
\begin{equation}\label{parity}
\mathcal{P}J_{\pm},\mathcal{P}=-J_{\pm}, \quad [h_D,\mathcal{P}]=[J_3,\mathcal{P}]=0.
\end{equation}
We will utilize these relations in the fourth section in the context of shape-invariance.
When $M=0$, the Hamiltonian $h_D$ also commutes with\footnote{Similar operator that 
switches the sign of coupling constant was employed in \cite{MikhailMirror} in the analysis of quantum systems with nonlinear supersymmetry.} 
$\mathcal{R}=R_{x_2}R_{c_3}\sigma_1$. The reflection operators $R_{x_2}$ 
and $R_{c_3}$ are defined as $R_{x_2}f(x_1,x_2,c_3)=f(x_1,-
x_2,c_3)$ and $R_{c_3}f(x_1,x_2,c_3)=$ $f(x_1,x_2,-c_3)$. For the generators of 
the potential algebra, there holds  
$$
\mathcal{R}J_{-}\mathcal{R}=J_+,\quad  \mathcal{R}J_{3}\mathcal{R}=-J_3.
$$

In our analysis, we will focus on the representations that can be 
constructed  from the lowest (highest) weight vectors, the zero-modes of 
ladder operator $J_-$ (or $J_+$). These vectors, denoted as  
$\psi_{m}^{-}$ (or $\psi_{m}^{+}$), can be fixed as the mutual eigenstates 
of $h_D$ and $J_3$,
\begin{equation}\label{lowesthighestconditions}
 J_{\pm}\psi_{m}^{\pm}=0, \quad J_{3}\psi_{m}^{\pm}=m\psi_{m}^{\pm}, \quad h_D\psi_{m}^{\pm}=E_{m}^{\pm}\psi_{m}^{\pm}.
\end{equation}
The energies $E^{\pm}_{m}$ can be 
found with the use of the relations (\ref{hDsquare}),
\begin{equation}\label{stacso3}
(E_{m}^{\pm})^2=\left\{\begin{array}{rl}-c_1\left(\left(m\pm\frac{1}{2}\right)^2-c_3^2\right)+M^2& \quad \mbox{for}\quad c_1\neq 0,\\4c_3\left(m\pm \frac{1}{2}\right) +M^2&\quad \mbox{for}\quad c_1=0.\end{array}\right.
\end{equation}
To keep the energies real, there must hold $\left|m\pm\frac{1}{2}\right|
\geq c_3$ for $c_1=-1$. For  $c_1=1$,  the values of $m$ are constrained by the strength of the coupling parameter, $\left|m\pm\frac{1}{2}\right|
\leq c_3$.
 
The other eigenstates of $h_D$ corresponding to the same energy can be 
found by repeated action of the ladder operator $J_+$ ($J_-$) on the vectors 
$\psi_{m}^-$ (or $\psi_{m}^+$). These eigenstates then establish irreducible representations of 
(\ref{algebra1})-(\ref{algebracentralelement}) 
that are specified by the energies $E_{m}^{\pm}$ of the corresponding lowest (highest) 
weight vector. 
These representations are usually 
associated with the bound states of the quantum system, see e.g. 
\cite{AdS2}. 

The unitary representations of $so(3)$ are finite dimensional.  
Consequently, the lowest weight vector, let us fix it as $\psi_{-m}^-$, has be to annihilated by 
specific power of the ladder operator $J_+$.  There holds  
$(J_+)^{d}\psi_{-m}^-=0$ where $d$ denotes dimension of the corresponding 
representation. The quantum number $m$ has to be either integer or semi-integer; we have $\psi_{m}^+\sim (J_{+})^{d-1}\psi_{-m}^{-}$ so that 
$J_3J_{+}^{d-1}\psi_{-m}^{-}=mJ_{+}^{d-1}\psi_{-m}^{-}$. Additionally, we can use (\ref{algebra1})
and write $J_3J_{+}^{d-1}\psi_{-m}^{-}=(-m+d-1)J_{+}^{d-1}\psi_{-m}^{-}$. 
Combining these relations together, there must hold $-2m+d-1=0$. 
Since $d$ is a positive integer, we get that $m$ has to be 
non-negative integer or semi-integer. 

The admissible values of the coupling constant $c_3$ depend on $m$; when 
$m$ acquires integer (half-integer) values, $c_3$ has to be half-integer 
(integer). We refer to Appendix A for details of the proof. Let us notice 
in this context that the representations of odd dimension for two 
different half-integer values of $c_3$ were 
considered in \cite{Vozmediano}. Representations of $so(3)$ and 
$so(2,1)$ were also discussed in \cite{Fakhri} where they were obtained 
analytically with the use of the master equation. 

For the purposes of the forthcoming section, let us discuss shortly how the results
can be used in description of the spin$-1/2$ particle  living in the two-dimensional curved space.
Let us consider 
the space where the metric tensor acquires diagonal form and the 
non-vanishing components depend on one coordinate only,
\begin{equation}\label{metric}
 g_{\mu\nu}=\left(\begin{array}{cc}g_{11}(x_1)&0\\0&g_{22}(x_1)\end{array}\right).
\end{equation}
In the most of the text, we will deal with the systems where both $x_1$ and $x_2$ are 
considered to be space-like ($g_{11}>0,$ $g_{22}>0$), i.e. the metric tensor corresponds to a curved 
surface embedded into three-dimensional euclidean space. Quantum settings of this kind appear frequently in the 
analysis of carbon nanostructures where deformations of graphene crystal 
are considered \cite{gaushrb}. The Dirac Hamiltonian can be written 
in the following non-covariant but manifestly Hermitian form (see Appendix B for details),
\begin{equation}\label{hDg}\nonumber
h_D=i\sigma_1\frac{1}
{g_{11}^{1/4}}\partial_{x_1}\frac{1}
{g_{11}^{1/4}}+\sigma_2\frac{J_3}{\sqrt{g_{22}}}+g_3\sigma_2+M\sigma_3.
\end{equation}
This operator is manifestly Hermitian with respect to the standard scalar product. 

When we compare (\ref{hDg}) with (\ref{hD}), it differs by the term containing $\partial_{x_1}$.  However, we can reduced (\ref{hDg}) into (\ref{hD}) by an appropriate change of coordinates. Let us set $z= z(x_1)$ such that $z'(x_1)=\sqrt{g_{11}}$. Rewriting (\ref{hDg}) in the new coordinate $z$, the multiplicative factor $g_{11}^{-1/2}$ in front of the derivative is eliminated. Then we make the similarity transformation $g_{11}^{-1/4}(x_1(z))\, h_D(\partial_{z},z,\partial_{x_2})\,g_{11}^{1/4}(x_1(z))$ which brings the operator back to the manifestly Hermitian form. In this way, the term $\frac{1}
{g_{11}^{1/4}}\partial_{x_1}\frac{1}
{g_{11}^{1/4}}$ in (\ref{hDg}) is effectively replaced by $\partial_{z}$. This transformed operator can be then identified with (\ref{hD}) for $g_2=g_{22}^{-1/2}(x_1(z))$ and $g_3=g_3(x_1(z))$.  

Using the inverse transformation, we can derive integrals of motion of (\ref{hDg}) from (\ref{Jpm}) that close the Lie algebra (\ref{algebra1})-(\ref{algebracentralelement})
\footnote{Generic form of the operators that can be obtained from (\ref{hD}) and (\ref{Jpm}) by the described transformation is
\begin{equation}\nonumber
 h_D=i\sigma_1\left(g_1\partial_{x_1}+\frac{g_1'}{2}\right)+\sigma_2\left(-ig_2\partial_{x_2}+g_3\right)+M\sigma_3, \quad g_a=g_a(x_1),\quad a=1,2,3,
\end{equation}
and 
\begin{equation}
 J_{\pm}=ie^{\pm ix_2}\left(g_1\partial_{x_1}+\frac{g_1'}{2}+\frac{g_1g_2'(\pm J_3+\frac{1}{2})}{g_2}\pm\frac{g_1g_3'}{g_2}\mp\frac{g_2}{2}\sigma_3\right).
\end{equation}
The functions $g_1$, $g_2$ and $g_3$ satisfy 
\begin{equation}\label{g1}
 g_1^2=\frac{g_2^4+c_1g_2^2}{(g_2')^2},\quad 
\quad g_3^2=\left\{\begin{array}{l}c_3^2(g_{2}^2+c_1),
\quad \mbox{for}\quad c_1\neq0,\\\\
c_3^2g_{2}^{-2},\quad \mbox{for}\quad c_1=0.\end{array}\right.
\end{equation}
Then these operators close (\ref{algebra1})-(\ref{algebracentralelement}).
 }. They are of the following form
\begin{eqnarray}\label{J}\nonumber
 J_{\pm}&=&i e^{\pm i x_2}\left(
\frac{1}
{g_{11}^{1/4}}\partial_{x_1}\frac{1}
{g_{11}^{1/4}}-\frac{g_{22}'}{2g_{22}\sqrt{g_{11}}}\left(\pm J_3+\frac{1}{2}\right)\pm \frac{\sqrt{g_{22}}g_3'}{ \sqrt{g_{11}}}
\mp\frac{1}{2\sqrt{g_{22}}}\sigma_3
\right),\quad J_3=-i\partial_{x_2},
\end{eqnarray}
provided that $g_{11}$, $g_{22}$ and the external potential $g_3$ satisfy the following equations
\begin{equation}\label{g1g}
 g_{11}=\frac{1}{4}\frac{(g_{22}')^2}{g_{22}(1+c_1g_{22})},\quad g_3=\left\{\begin{array}{l}c_3\,\sqrt{\frac{1}{g_{22}}+c_1},\quad \mbox{for}\quad c_1\neq0,\\\\
                                                     c_3\,g_{22}^{1/2},\quad \mbox{for}\quad c_1=0.\end{array}\right.
\end{equation}
We will use the formulas (\ref{hDg}), (\ref{J}) and (\ref{g1g}) extensively in the forthcoming text.

Concluding the section, let us notice that the operator (\ref{hDg}) can describe low-energy charge carriers in two-dimensional crystals with hexagonal lattice where the (nontrivial) geometry of the crystal surface is encoded in $g_{\mu\nu}$. In particular, we refer to graphene and boron-nitride crystals for $M=0$ and $M\neq0$ respectively, see \cite{Semenoff}.  In this context, the generally non-constant coefficients of $\partial_{x_1}$ and $\partial_{x_2}$ in (\ref{hDg}) can be interpreted as the space-dependent modulation of Fermi velocity of massless Dirac particle in graphene induced by the strains and ripples of the crystal \cite{Fermimodulation}. The non-vanishing vector potential represented by $g_3$ can be induced by external magnetic fields or by defects of the crystal lattice \cite{Vozmediano}, \cite{VozmedianoGauge}.

\section{Solvable models of open-cage fullerenes\label{carved}}
Fullerenes are spherical molecules made of carbon atoms. The spherical 
shape is due to twelve pentagons that are inserted into hexagonal carbon 
lattice. The most famous example is the $C_{60}$, ``Buckminsterfullerene'' 
\cite{Buckminster}, which can be classified as a truncated icosahedron 
with 60 vortices and 32 faces.  Fixing the number of pentagons in the 
lattice while increasing the number of hexagons, we can get not only 
bigger fullerenes like $C_{240}$ or $C_{540}$, but also objects of more 
complicated shapes, e.g. capped nanotubes, elliptic fullerenes etc.. 
We will be interested just in the systems where the surface of 
the crystal can be approximated by the spherical geometry.

In quantum chemistry, the situation is considered where small enough atoms 
or molecules (e.g. of hydrogen) are inserted inside the fullerenes. During 
the process called ``chemical surgery'', some bonds between the carbon 
atoms of the fullerene are broken, making the hole into the surface. The 
alien atoms are then inserted into the opened cage and the hole is closed 
again \cite{OpenCage}. In this section, we will focus on the electronic 
properties of the open-cage fullerenes with one or two holes in the 
surface. We will construct exactly solvable models that can serve as the 
low-energy approximation of these systems. 

\subsection{Spectrum of fullerenes}
First, let us go briefly through the analysis of the electronic properties 
of the (unopened) fullerenes. The low-energy excitations of charge 
carriers in the crystal can be investigated efficiently within the 
framework of the Dirac equation on the spherical surface 
\cite{Vozmediano}. The surface embedded into three-dimensional space can 
be parametrized  in spherical coordinates as
$x=\sin x_1\cos x_2$, $y=\sin x_1\sin x_2$, $z=\cos x_1$ where 
$x_1\in \langle0,\pi\rangle$ and $x_2=\langle 0,2\pi\rangle$. 
The associated metric tensor is explicitly
\begin{equation}\label{lenghtonehole}\nonumber
 g_{\mu\nu}=\left(\begin{array}{cc}1&0\\0&\sin^2 x_1\end{array}\right).
\end{equation} 
The entries of the metric tensor satisfy (\ref{g1g}) for $c_1=-1$. 
Employing the formula (\ref{hDg}), we can write down immediately the 
Dirac Hamiltonian corresponding to the massless, spin$-1/2$ particle on the sphere,
\begin{equation}\label{hDsphere}
 h_D=i\sigma_1\partial_{x_1}+\sigma_2\frac{J_3}{\sin x_1}+c_3\sigma_2\cot x_1,\quad x_1\in(0,\pi).
\end{equation}
The potential term corresponds to Dirac monopole that is situated in the center of the sphere. 
It occurs due to the topological defects, twelve pentagons, that have to 
be inserted into the hexagonal lattice of graphene to close the spherical 
surface \cite{Osipov},  \cite{VozmedianoGauge}, \cite{Vozmediano}. 
The actual value of the coupling constant is  $c_3=\frac{n}{8}$ where $n$
corresponds to the number of conical defects in the crystal. As $n=12$ for fullerenes,
we get $c_3=\frac{3}{2}$, see  \cite{Osipov}, \cite{Vozmediano}.

We will not write down the explicit form of $J_{\pm}$, it can be 
extracted easily from (\ref{J}). The lowest weight vectors can be found explicitly by solving  
(\ref{lowesthighestconditions}). They are
\begin{eqnarray}\label{lowestfulleren}
 \psi_{m}^-&=&e^{i m x_2}w_F^{-c_3} \left(\sin x_1\right)^{-m+\frac{1}{2}} \left(\begin{array}{r}\sqrt{m-\frac{1}{2}+c_3}\, w_F^{-\frac{1}{2}}\\-\sqrt{-m+\frac{1}{2}+c_3}\,w_F^{\frac{1}{2}}\end{array}
 \right),\quad w_F=\tan\frac{x_1}{2},\\
 h_D\psi_{m}^-&=&\sqrt{\left(m-\frac{1}{2}\right)^2-c_3^2}\,\psi_{m}^-.
\end{eqnarray}
We require the eigenfunctions of (\ref{hDsphere}) to be square integrable 
and vanishing at $x_1=0,\pi$. 
The vectors (\ref{lowestfulleren}) are physically acceptable as long as 
$m\leq -|c_3|+\frac{1}{2}$. In that case, they are regular and correspond 
to real energies, see the restriction below (\ref{stacso3}). Degeneracy of 
the zero-energy level is equal to the dimension of the representation of 
$so(3)$ spanned by zero modes of the Hamiltonian. For $c_3=\frac{3}{2}$,
the zero-energy has triple degeneracy.

For purposes of the forthcoming analysis, it is convenient to define 
transformation $G_F$,
\begin{equation}\nonumber
 G_F=w_F^{c_3}\sqrt{\sin x_1}\left(\begin{array}{cc}w_F^{1/2}&0\\0&w_F^{-1/2}\end{array}\right)\sigma_1,
\end{equation}
and rewrite the lowest weight vector in the factorized form 
\begin{equation}\label{gaugedpsikoule}
 \psi_{m}^-=e^{imx_2}\,y^{-m}G_F\left(\begin{array}{r}-\sqrt{-m+\frac{1}{2}+c_3}\, w_F^{k_{u}}\\\sqrt{m-\frac{1}{2}+c_3}\, w_F^{k_{d}}\end{array}\right),\quad y=\sin x_1,\quad w_F(y)=\left\{^{\frac{1-\sqrt{1-y^2}}{y},\ x_1<\pi/2,}_{\frac{1+\sqrt{1-y^2}}{y},\ x_1\geq\pi/2,}\right.
\end{equation}
where 
\begin{equation}\label{kukd} 
k_u=1-2c_3,\quad k_d=-2c_3-1.
\end{equation}
With the use of this notation, the ladder operator can be rewritten as
\begin{eqnarray}\label{gaugedJkoule}
 G^{-1}J_+G&=&ie^{ix_2}\left(\partial_{x_1}-\cot x_1 J_3\right)=\quad ie^{ix_2}(1-y\,w_F(y))\left(\partial_{y}-\frac{J_3}{y}\right).
\end{eqnarray}
It would be necessary to check that a multiple action of $J_+$ on 
$\psi_{m}^-$ does not violate regularity of the wave functions. However, 
we can skip this task as the considered wave functions are known to be 
regular and based on the Jacobi polynomials, see 
\cite{Fakhri}, \cite{Campoersi}. It means that the action of (\ref{gaugedJkoule}) 
on (\ref{gaugedpsikoule}) preserves the required boundary conditions. We 
will use this fact in the analysis of the open-cage fullerenes. 

\subsection{Open-cage fullerene with one hole}
Let us consider the fullerene molecule with one hole carved into its 
surface (see Fig.1). We suppose that the twelve pentagons allowing the 
crystal to close the spherical shell are still present in the lattice. To 
meet this requirement, we can suppose that the model describes larger 
molecules like $C_{240}$ where one or few hexagons can be extracted from 
the crystal without affecting the pentagons. The considered spherical 
surface can be parametrized as 
\begin{equation}\nonumber
 x=\sin \left(\pi dn(x_1,k)\right)\cos x_2,\quad 
y=\sin \left(\pi dn(x_1,k)\right)\sin x_2\quad z=\cos \left(\pi dn(x_1,k)\right),
\end{equation}
where $x_1\in\langle 0,K\rangle,\ x_2=\langle -\pi,\pi\rangle$ and $k\in (0,1)$.
The components of the associated metric tensor 
\begin{equation}\nonumber
 g_{\mu\nu}=\left(\begin{array}{cc}\left(\pi k^2 cn(x_1,k)sn(x_1,k)\right)^2&0\\
0&\sin^2\left(\pi dn(x_1,k)\right)\end{array}\right)
\end{equation}
satisfy (\ref{g1g}) for $c_1=-1$.

The functions $dn(x_1,k)$, $cn(x_1,k)$ and $sn(x_1,k)$ are Jacobi elliptic 
functions and $K=\int^{\frac{\pi}{2}}_{0}(1-k^2sin^2 \theta)^{-1}\,d\theta$ 
is the complete elliptic integral of the first kind with $0\leq k\leq 1$. 
Let us briefly refresh relevant properties of these functions. $dn(x_1,k)$ is periodic with periodicity $2K$ while $cn(x_1,k)$ and $sn(x_1,k)$ have periodicity 
$4K$. There holds $0<dn(x_1,k)\leq 1$ for all real $x_1$ and all values of the modular parameter $k\in \langle 0,1\rangle$.  The functions $cn(x_1,k)$ and $sn(x_1,k)$ can be understood as a 
generalization of trigonometric functions; they 
satisfy the relation $cn(x_1,k)^2+sn(x_1,k)^2=1$.   There holds 
$cn(x_1,0)=\cos x_1$ and $sn(x_1,0)=\sin x_1$, $dn(x_1,0)=1$. When $k\rightarrow 1$, the period of the Jacobi elliptic functions tends to 
infinity. There holds $sn(x_1,1)=\mbox{tanh} x_1 $, whereas 
$sn(x_1,1)=dn(x_1,1)=\mbox{sech} x_1$. In the cases where the actual value 
of the modular parameter is not informative or it is clear from the 
context, we will use the shortened notation $sn(x_1)\equiv sn(x_1,k)$, 
$cn(x_1)\equiv cn(x_1,k)$ and $dn(x_1)\equiv dn(x_1,k)$.  

Dirac Hamiltonian on the considered surface acquires the following (non-covariant but manifestly Hermitian) 
explicit form
\begin{equation}\label{hDonehole}
 h_D=i\sigma_1\frac{1}{k^2\pi cn x_1 sn x_1}\left(\partial_{x_1}+\frac{dn x_1(sn^2x_1-cn^2x_1)}{2 cn x_1 sn x_1}\right)+\sigma_2\frac{J_3}{\sin(\pi dn x_1)}+c_3\sigma_2\cot (\pi dn x_1),
\end{equation}
where $x_1\in (0,K)$.
Likewise in (\ref{hDsphere}), the potential term corresponds to the field 
generated by Dirac monopole. We will require the eigenstates of $h_D$ to 
be regular and vanishing at the edge of the hole (i.e. at $x_1=K$). 
The lowest weight vectors $\psi_{m}^-$ annihilated by $J_-$ are
\begin{eqnarray}\label{onehole}
 \psi_{m}^-&=&e^{im x_2}\frac{\sqrt{cn x_1 sn x_1}}{w_I^{c_3}}(\sin(\pi dn x_1))^{-m+\frac{1}{2}}
\left(\begin{array}{l}
\sqrt{c_3+\frac{1}{2}-m}\, w_I^{1/2} 
\\ 
\sqrt{c_3-\frac{1}{2}+m}\,w_I^{-1/2}\end{array}\right),\end{eqnarray} where $w_I=\tan\frac{\pi dn x_1}{2}$ and $
h_D\psi_{m}^-=\sqrt{(m-1/2)^2-c_3^2}\,\psi_{m}^-.$

The wave functions (\ref{onehole}) comply with the required boundary 
condition $(cn(K)=0)$. However, it is not clear 
whether the multiple action of $J_+$ on (\ref{onehole}) does not violate 
the  boundary conditions. We can 
analyze this point using the close analogy with the model of fullerenes. 
Let us define the transformation $G_I$ as  
\begin{equation}\nonumber
G_I= \sqrt{cn x_1 sn x_1 \sin(\pi dn(x_1))}\,w_I^{c_3}\left(\begin{array}{cc}w_I^{-1/2}&0\\0&w_I^{1/2}\end{array}\right).
\end{equation}
The lowest weight vector can be then written as
\begin{equation}\label{gaugedpsionehole}
 \psi_{m}^-=e^{imx_2}y^{-m}G_I\left(\begin{array}{l}\sqrt{c_3+\frac{1}{2}-m}\, w_I^{k_{u}}\\\sqrt{c_3-\frac{1}{2}+m}\,w_I^{k_{d}}\end{array}\right),
\quad y=\sin (\pi dn x_1),
\quad w_I(y)=\left\{^{\frac{1-\sqrt{1-y^2}}{y},\ dn x_1\leq 1/2,}_{\frac{1+\sqrt{1-y^2}}{y},\  dn x_1>1/2,}\right.
\end{equation}
where the constants $k_u$ and $k_d$ are those in (\ref{kukd}). 
Performing the ``gauge'' transformation of $J_+$, we get
\begin{eqnarray}\label{gaugedJ}
 G_I^{-1}J_+G_I&=&ie^{ix_2}\left(\frac{1}{k^2\pi cn x_1 sn x_1}\partial_{x_1}+\cot \pi dn x_1 J_3\right)=\quad -ie^{ix_2}(1-y\, w_I(y))\left(\partial_{y}-\frac{J_3}{y}\right).
\end{eqnarray}
The latter operator is identical with (\ref{gaugedJkoule}). The vector 
(\ref{gaugedpsionehole}) coincides with (\ref{gaugedpsikoule}) 
up to the factors $G_F$ and $G_I$ and up to the
sign of the upper component. Hence, likewise in the case of fullerenes, 
the multiple action of $J_+$ on $\psi_{m}^-$ will keep regularity of the 
wave functions, i.e. only positive powers of $y$ will emerge. 

\subsection{Open-cage fullerene with two holes}
Similar results can be obtained for spin-$1/2$ particle on the spherical 
surface with two holes centered in the poles, see Fig.1. The 
surface can be parametrized as
$
 x= dn x_1\,\cos x_2$, $y=dn x_1\,\sin x_2,$ $
z=k sn x_1$, where $x_1\in\langle-K,K\rangle$, $x_2\in\langle -\pi,\pi\rangle
$ and $k\in (0,1)$.
The metric tensor reads this time
\begin{equation}\nonumber
g_{\mu\nu}=\left(\begin{array}{cc}\left(k cn x_1\right)^2&0\\0&dn^2 x_1\end{array}\right).
\end{equation}
The explicit form of the Hamiltonian can be written as follows
\begin{equation}\label{hDtwoholes}
 h_D=\frac{i\sigma_1}{kcn x_1}\left(\partial_{x_1}+\frac{dn x_1 sn x_1}{2cn x_1}\right)+\sigma_2\frac{J_3}{dn x_1}+c_3\sigma_2\frac{k sn x_1}{dn x_1},\quad x_1\in(-K,K).
\end{equation}
\newsavebox{\onehole}        
    \savebox{\onehole}{          
    \scalebox{1}{
    \includegraphics[scale=.4]{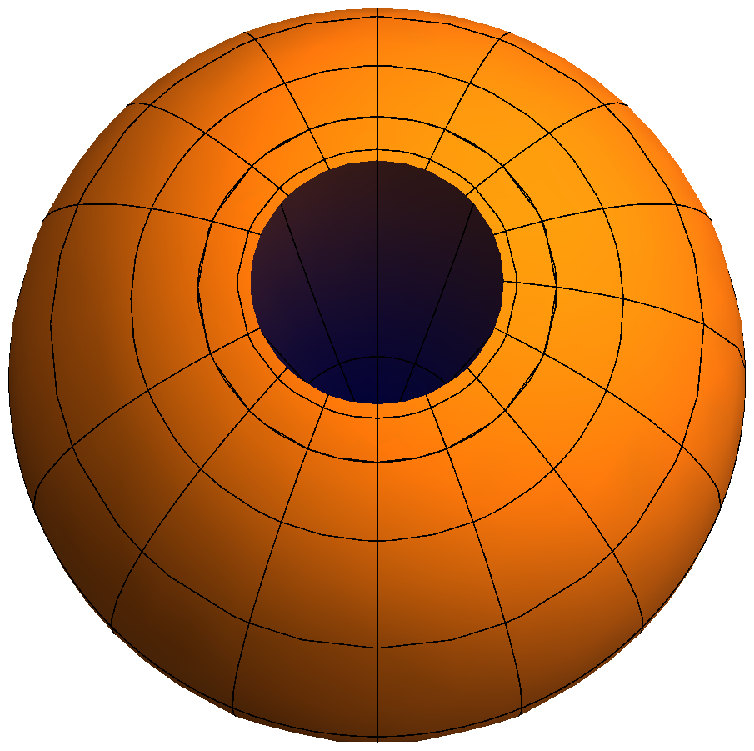}
    }
    }

\newsavebox{\twoholes}        
    \savebox{\twoholes}{          
    \scalebox{1}{
    \includegraphics[scale=.4]{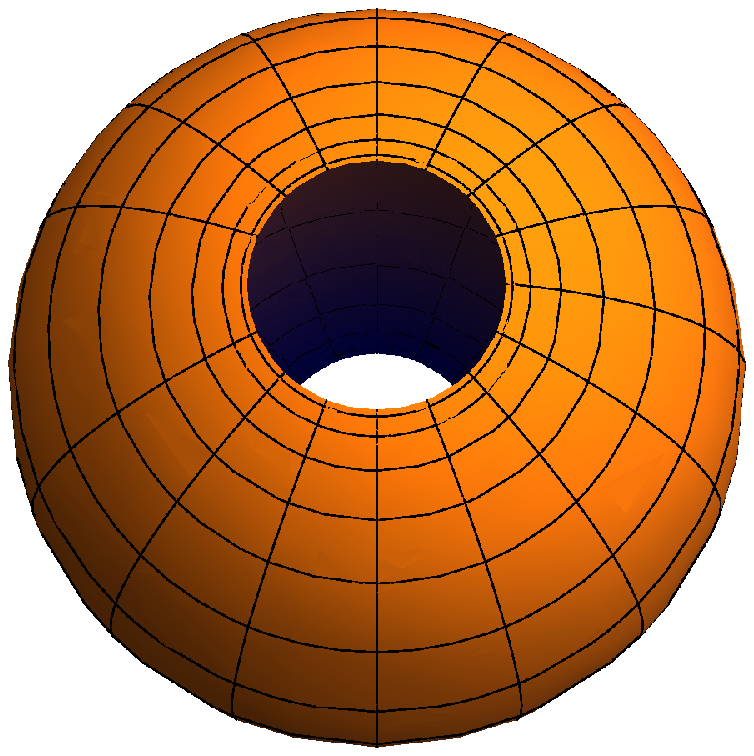}
    }
    }

\begin{figure}[h!]\begin{center}
\usebox{\onehole}\usebox{\twoholes}\caption{The models of open cage fullerenes. We fixed $k=.99$ in the left figure while $k=.9$ in the right figure.}
\end{center}\label{figfuller}\end{figure}
\noindent
Its eigenvectors should be regular and vanishing at the edges of the holes 
(i.e. for $x_1\rightarrow \pm K$). The lowest weight vector $\psi_{m}^-$ 
which satisfies $h_D\psi_{m}^-=\sqrt{(m-\frac{1}{2})^2-c_3^2}\psi_{m}^-$ 
has the explicit form  
\begin{eqnarray}\label{twoholes}
 \psi_{m}^-&=&\frac{\sqrt{cn x_1 }}{w_{II}^{c_3}}(dn x_1)^{-m+\frac{1}{2}}
\left(\begin{array}{l}
\sqrt{c_3+\frac{1}{2}-m}\, w_{II}^{1/2} 
\\ 
\sqrt{c_3-\frac{1}{2}+m}\,w_{II}^{-1/2}\end{array}\right),\quad w_{II}=\left(\frac{1-k sn x_1}{dn x_1}\right).
\end{eqnarray}
It complies with the required boundary conditions since 
$cn( K)=cn (-K)=0$.
In order to show that application of $J_+$ on these vectors does not 
violate the required properties, let us define the matrix
\begin{equation}\nonumber
G_{II}=\sqrt{cn(x_1)dn(x_1)}\,w_{II}^{c_3}\left(\begin{array}{cc}w_{II}^{-1/2}&0\\0&w_{II}^{1/2}\end{array}\right).
\end{equation}
The lowest weight vector can be written in the following form
\begin{eqnarray}\label{lowestsud}
\psi_{m}^-&=&e^{im x_2}\, y^{-m} G_{II}\left(\begin{array}{c}\sqrt{c_3+\frac{1}{2}-m}\, w_{II}^{k_u}\\\sqrt{c_3-\frac{1}{2}+m}\,w_{II}^{k_d}\end{array}\right),
\quad y=dn x_1,
\quad w_{II}(y)=
\left\{^{\frac{1-\sqrt{1-y^2}}{y},\ sn x_1> 0,}_{\frac{1+\sqrt{1-y^2}}{y},\  sn x_1\leq0,}\right.
\end{eqnarray}
while the ladder operator is transformed into
\begin{equation}\label{gaugedJtwohole}
 G_{II}^{-1} J_+ G_{II}=i e^{ix_2}\left(\frac{1}{k cn x_1}\partial_{x_1}+\frac{J_3 k sn x_1}{dn x_1}\right)=-i e^{ix_2}(1-y w_{II}(y))\left(\partial_{y}-\frac{J_3}{y}\right).
\end{equation}
Comparing (\ref{lowestsud}) with (\ref{gaugedpsikoule}) and 
(\ref{gaugedJtwohole}) with (\ref{gaugedJkoule}), we can conclude 
that the repeated action of $J_+$ on $\psi_{m}^-$ will preserve 
regularity of the wave functions. 

The formulas (\ref{gaugedpsionehole}), (\ref{gaugedJ}), (\ref{lowestsud}) 
and (\ref{gaugedJtwohole}) allow to find the required eigenstates 
of the corresponding Hamiltonian as well as energies which are given by 
(\ref{stacso3}) for $c_1=-1$.  Each of the systems has  distinct 
topological genus as the number of the holes in the spherical surface 
varies. Despite this distinct topological nature, the models are spectrally identical and possess the same algebraic background. This follows from the fact that they can be related by the coordinate transformation discussed in the preceding section.

In the next section, we will discuss a different application of the 
potential algebras. We will consider one-dimensional, both relativistic and non-relativistic supersymmetric systems that possess shape invariance. 

\section{Shape invariant Dirac-like Hamiltonians and associated nonrelativistic systems}
The concept of shape invariance was proposed by Gendenshtein \cite{Gendenstein}. It originated from the supersymmetric quantum mechanics dominated by Witten's model \cite{Witten}, 
 \cite{Khare}. It attracted a lot of attention for being extremely useful in construction and analysis of exactly solvable models. Instead of going into the details of the concept, let us sketch briefly the main idea that will be relevant for our forthcoming discussion. 

Let us suppose that the Hamiltonian $H_m$, describing a (non-relativistic) quantum system, contains a potential term that depends explicitly on the coupling constant $m$. The ground state $\psi^{(0)}_m$ of $H_m$ is known for any value of the coupling constant $m$ and corresponds to the ground state energy $E^{(0)}_m$, $H_m\psi^{(0)}_m=E_{m}^{(0)}\psi^{(0)}_m$. Suppose that the ground state energy is a decreasing function of the coupling constant, i.e. $E^{(0)}_{n}>E^{(0)}_m $ for $n<m$.  Finally, let us take for granted that there exist operators $A_m$ and $A_m^{\dagger}$ which intertwine $H_m$ with another Hamiltonian $\tilde{H}$ that differs from $H_m$ just by the value of the coupling constant, i.e. $\tilde{H}=H_{m-\delta}$. There holds
\begin{equation}\label{SIP}
 A_mH_m=H_{m-\delta}A_m,\quad H_mA^{\dagger}_m=A^{\dagger}_mH_{m-\delta}.
\end{equation}
Then we say that the $H_m$ is shape-invariant.
For such setting, we can easily obtain excited states with energy $E_{m}^{(n)}$. Indeed, it is sufficient to apply appropriate intertwining operators on the ground state $\psi_{m-n\delta}^{(0)}$ of $H_{m-n\delta}$. We get the following formula,
\begin{equation}\label{SIPstates}
 (H_{m}-E_{m}^{(n)})A^{\dagger}_{m}A^{\dagger}_{m-\delta}..A^{\dagger}_{m-(n-1)\delta}\psi_{m-n\delta}^{(0)}=0,\quad E_{m}^{(n)}=E_{m-n\delta}^{(0)}.
\end{equation}
In should be kept in mind that when a specific system is considered, one has to check that the intertwining operators do not violate boundary conditions prescribed for the eigenfunctions, that they preserve domains of the Hamiltonians etc..  

As an example of non-relativistic shape-invariant setting, let us mention Poschl-Teller system described by the Hamiltonian $H_m=-\partial_{x_1}^2-\frac{m(m+1)}{\cosh^2x_1}$. It has the ground state energy $E^{(0)}_{m}=-m^2$ and the ground state $\psi_m^{(0)}=\cosh^{-m}x_1$. There holds $A_mH_{m}=H_{m-1}A_m$ where $A_{m}=\partial_{x_1}+m\tanh x_1$. It can be identified with (\ref{SIP}) for $\delta=1$. The explicit form and the energies of the excited bound state can be found directly by the formula (\ref{SIPstates}). We have $E^{(n)}_m=-(m-n)^2$, $n=0,1,\dots,m-1$. 

The non-relativistic systems with shape-invariant potentials were studied extensively in the literature, see e.g. \cite{Levai}, \cite{Nikitin}. The systematic analysis was possible as the  (second-order) Hamiltonians could be factorized in terms of the first order differential operators.
The situation gets more complicated in case of Dirac operators which are itself of the first order, and, hence, cannot be factorized in the similar way. 

Usually, the shape-invariance of Dirac Hamiltonian is 
treated such that the square of the operator is identified with a 
known, nonrelativistic, shape-invariant operator after a series of 
transformations \cite{Alhaidari}. In this manner, exact eigenstates and 
eigenvalues of the original relativistic system can be found. 
Instead of following this way, we will consider directly the shape-invariance of the \textit{first-order} Dirac-like Hamiltonians. In our approach, we will be inspired by Balantekin and 
Gangopadhyaya \cite{Balantekin}, \cite{Gango}. They found that the shape-invariance of  (nonrelativistic) systems can be understood in terms of a 
higher dimensional Hamiltonian that possesses Lie algebra of integrals of 
motion. In this framework, the Hamiltonian $H_{m}$ as 
well as the $A_m$ (and $A_m^{\dagger}$) 
correspond to an appropriate restriction of the higher-dimensional 
Hamiltonian and of the ladder 
operators to the subspaces with fixed value of angular momentum. 

\subsection{Shape-invariance of Dirac operators via $N=4$ nonlinear supersymmetry}
Let us consider restriction of the Hamiltonian (\ref{hD}) and of the ladder 
operators (\ref{Jpm}) to the subspaces where $J_3$ acquires 
fixed value.
The operator $h_D$ reduces to the one-dimensional Dirac Hamiltonian
\begin{equation}\label{hrestricted}
h_m=h_D|_{J_3= m}=e^{-imx_2}h_De^{imx_2}=i\sigma_1\partial_{x_1}+\sigma_2g_2m+g_3\sigma_2+M\sigma_3.
\end{equation}
The ladder operators $J_{\pm}$ exchange the subspaces where 
$J_3= m$ and $J_{3}= m+1$.
The restricted operators $j_{\pm,m}$ read explicitly 
\begin{eqnarray}\label{j}
 j_{m}^+&=&e^{-i(m+1)x_2}J_+e^{imx_2}
 =i\left(\partial_{x_1}+\frac{g_2'(m+\frac{1}{2})+g_3'}{g_2}-\frac{g_2}{2}\sigma_3\right),\nonumber\\  
 j_{m}^-&=&e^{-imx_2}J_-e^{i(m+1)x_2}
=i\left(\partial_{x_1}-\frac{g_2'(m+\frac{1}{2})+g_3'}{g_2}+\frac{g_2}{2}\sigma_3\right).
\end{eqnarray}
Keeping in mind that  $[h_D,J_{\pm}]=0$, we can write down the following 
intertwining relations between the operators $h_m$ and $h_{m+1}$,
\begin{equation}\label{DiracSIP}
h_mj_{m}^-=j_{m}^-h_{m+1},\quad  h_{m+1}j_{m}^+=j_{m}^+h_{m}.
\end{equation}
Comparing with (\ref{SIP}), they establish the shape-invariance of the one-dimensional Dirac operator $h_m$ mediated by the operators $j_{m}^{\pm}$. 
The operators $j_{m}^{\pm}$ can be identified as the generalized matrix Darboux 
transformation. General properties of these transformations were discussed in \cite{DiracDarboux} in 
detail. In particular, it was proved  that the product $j_{m}^-j_{m}^+$ or 
$j_{m}^+j_{m}^-$ is proportional to the second-order polynomials in $h_m$ 
or $h_{m+1}$ respectively, 
\begin{equation}\label{shapeDirac}
 j_{m}^-j_{m}^+=h_m^2-\mathcal{E}_m,\quad j_{m}^+j_{m}^-=h_{m+1}^2-\mathcal{E}_m,
\end{equation}
where  $\mathcal{E}_m=(E^+_m)^2$, see (\ref{stacso3}). Let us notice that similar approach to the shape-invariance of Dirac 
operators based on the intertwining relations (\ref{DiracSIP}) also appeared recently in \cite{Negro}.

We can understand the relations (\ref{DiracSIP}) and (\ref{shapeDirac}) as the manifestation of a supersymmetry where $j_{m}^{\pm}$ are components of the supercharges while $h_m$ and $h_{m+1}$ compose the superextended Hamiltonian. Picking up $\tau_3$ as the grading operator, the operators 
\begin{eqnarray}\label{Diracshapeinvariance}
 &&
   \mathtt{h}_m=\left(\begin{array}{cc}h_m&0\\0&h_{m+1} \end{array}\right),\quad 
   \mathcal{J}_{m}^{(1)}=\left(\begin{array}{cc}0&j_{m}^-\\j_{m}^+&0 \end{array}\right),\quad 
   \mathcal{J}_{m}^{(2)}=i\, \tau_3 \mathcal{J}_{m}^{(1)},
\end{eqnarray}
form the $N=2$  supersymmetry, 
\begin{eqnarray}\label{dnsusy1}
&& [\mathtt{h}_m,\mathcal{J}_{m}^{(a)}]=0,\quad 
\{\mathcal{J}_{m}^{(a)},\mathcal{J}_{m}^{(b)}\}=2\delta_{ab}\left(\mathtt{h}_m^2-\mathcal{E}_m\right), \quad a,b=1,2.
\end{eqnarray}
The supersymmetry is 
nonlinear, see \cite{nSUSY}, as the anticommutator of the supercharges is a quadratic 
polynomial in the (extended) Hamiltonian $\mathtt{h}_m$.  The operators (\ref{Diracshapeinvariance}) act on the bispinors with the upper 
spinor from the subspace where $J_3= m$ and with the lower one from the 
subspace where $J_3= m+1$. 
After the restriction, the operator $J_3$ reduces to $ \mathcal{J}_{m}^{(3)}=\mbox{diag}(m,m+1)=\left(m+\frac{1}{2}\right)(\mathbf{1}\otimes\mathbf{1})-\frac{\tau_3}{2}$ where $\tau_3=\sigma_3\otimes\mathbf{1}$. The intertwining relations (\ref{DiracSIP}) of the relativistic Hamiltonian (\ref{hrestricted}) are encoded in the first commutator of (\ref{dnsusy1}).  

The choice the grading operator is not unique; we could equally well accept the parity $\mathcal{P}$ (see (\ref{parity})) in this role, $[\mathcal{P},\mathtt{h}_m]=\{\mathcal{J}^{(a)}_{m},\mathcal{P}\}=0$, and define $\mathcal{J}^{(2)}_m=i\mathcal{P} \mathcal{J}_{m}^{(1)}$. Then exactly the same superalgebra (\ref{dnsusy1}) would emerge (with distinct realization of the supercharges). To treat these two parallel algebraic structures in a unified framework, let us consider the following fermionic operators 
\begin{equation}\label{JJ}
 \mathcal{J}_m^{(1,1)}=\mathcal{J}_m^{(1)},\quad \mathcal{J}_m^{(2,1)}=i\, \mathcal{P} \mathcal{J}_{m}^{(1)},
\quad \mathcal{J}_m^{(1,2)}=\,\tau_3\, \mathcal{P} \mathcal{J}_{m}^{(1)},\quad \mathcal{J}_m^{(2,2)}=\,i\tau_3\, \mathcal{J}_{m}^{(1)}.
\end{equation}
They close the following  $N=4$ \textit{nonlinear} superalgebra where the bosonic operators $\mathcal{P}$ and $\tau_3$ are included,
\begin{eqnarray}
 &&[\mathtt{h}_m,\mathcal{J}_{m}^{(a)}]=0,\quad \{\mathcal{J}_m^{(a,c)},\mathcal{J}_m^{(b,d)}\}=2\delta_{ab}(\delta_{cd}+(1-\delta_{cd})\tau_3\, \mathcal{P})(\mathtt{h}_m^2-\mathcal{E}_m),\\
 && [\mathcal{P},\mathcal{J}^{(a,b)}_m]=-2i\varepsilon_{ac}\mathcal{J}^{(c,b)}_m,\quad [\tau_3,\mathcal{J}^{(a,b)}_m]=-2i\left(\delta_{ab}\varepsilon_{ac}\mathcal{J}^{(c,c)}_m-\varepsilon_{ab}\varepsilon_{ac}\varepsilon_{bd}\mathcal{J}^{(c,d)}_m\right).
\end{eqnarray}
Here $\varepsilon_{ab}$ is completely antisymmetric in indices and 
$\varepsilon_{12}=1$. Hence, the shape invariance of $h_m$ is associated with nonlinear $N=4$ 
superalgebra.
 
For $M=0$, the operator $H_m=h_m^2$ coincides with the Hamiltonian of one of the nonrelativistic shape-invariant models that can be, dependently on the actual choice of $g_2$ and $g_3$ (see (\ref{g2})), classified as the trigonometric Scarf I (for $c_1=-1$), Harmonic oscillator ($c_1=0$) or Rosen-Morse II system ($c_1=1$), see \cite{DeDuttSukhatme}. In the regime of zero mass, the operator $\sigma_3$ anticommutes with $h_m$ and, hence, it can be regarded as the grading operator of the standard $N=2$ supersymmetry that underlies shape-invariance of $H_m$.  It reads explicitly $[H_m,Q_m^{(a)}]=0$, $\{Q_m^{(a)},Q_m^{(b)}\}=2\delta_{ab}H_m$, where $Q_m^{(1)}=h_m$, $Q_m^{(2)}=i\sigma_3h_m$ and  $a,b=1,2$.

Is it possible to extend this structure with the fermionic operators $\mathcal{J}^{(a,b)}_m$?
In order to do so, we define the 
following extended operators
\begin{equation}\label{PDMextended}
\mathcal{H}_m=\mathtt{h}_m^2=\mbox{diag}(H_m,H_{m+1}),\quad  \mathcal{Q}_{m}^{(1)}=\mathtt{h}_m,\quad  \mathcal{Q}_{m}^{(2)}=i\mathbf{1}\otimes\sigma_3\mathcal{Q}_{m}^{(1)}.
\end{equation} 
that close reducible $N=2$ superalgebra graded by $\mathbf{1}\otimes \sigma_3$. In order to treat both $\mathcal{Q}_{m}^{(a)}$ and $\mathcal{J}_m^{(a,b)}$ as fermionic operators, we have to fix $\sigma_3\otimes \sigma_3$ as the grading operator. There are sixteen fermionic operators in the extended superalgebra; half of them are local and half are nonlocal operators. They are $\mathcal{J}^{(a,b)}_m$ and $i\sigma_3\otimes\sigma_3\mathcal{J}^{(a,b)}_m$ together with $\mathcal{Q}_{m}^{(a)}$, $\tau_3\mathcal{P}\mathcal{Q}_{m}^{(a)}$, $i(\mathbf{1}\otimes\sigma_3)\mathcal{Q}_{m}^{(a)}$ and $i(\sigma_3\otimes\sigma_3)\mathcal{Q}_{m}^{(a)}$. Additionally, it is necessary to introduce bosonic operators 
$\mathcal{X}_{m}^{(a,b)}=\mathcal{J}_{m}^{(a)}\mathcal{Q}_{m}^{(b)}$ and $\mathcal{P}\mathcal{X}_{m}^{(a,b)}$, $a,b=1,2.$  Instead of writing down all the relation of the superalgebra, let us focus on its part generated by the local operators. 
Then we have just eight fermionic operators which we
denote in the following manner,  
\begin{equation}\label{fermionic}\nonumber
\mathcal{Q}_{m}^{(a,1)}\equiv\mathcal{Q}_{m}^{(a)},\quad \mathcal{Q}_{m}^{(a,2)}\equiv\tau_3\mathcal{Q}_{m}^{(a)}\quad \tilde{\mathcal{J}}_{m}^{(a,1)}\equiv \mathcal{J}_{m}^{(a)},\quad  \tilde{\mathcal{J}}_{m}^{(a,2)}\equiv\tilde{\sigma}_3\mathcal{J}_{m}^{(a)},\quad\tilde{\sigma_3}=\mathbf{1}\otimes\sigma_3.\end{equation}
The sub-superalgebra generated by local operators then reads\footnote{In the computation of the commutators and anticommutators, we used 
$[\mathcal{J}_{m}^{(a)},\mathcal{Q}_{m}^{(b)}]=0$, 
$[\mathcal{J}_{m}^{(a)},\mathcal{J}_{m}^{(b)}]=-2i\varepsilon_{ab}\tau_3(\mathcal{H}_m-\mathcal{E}_m)$, $[\mathcal{Q}_{m}^{(a)},\mathcal{Q}_{m}^{(b)}]=-2i\varepsilon_{ab}\tilde{\sigma}_3\mathcal{H}_m$. } 
\begin{eqnarray}\label{PDMWITTEN}
\ [\mathcal{H}_m,\mathcal{Q}_m^{(a,b)}]&=&0, \quad \{\mathcal{Q}_m^{(a,c)},\mathcal{Q}_m^{(b,d)}\}=2\delta_{ab}(\delta_{cd}+(1-\delta_{cd})\tau_3)\mathcal{H}_m,\\
\label{PDMSHAPE} [\mathcal{H}_m,\tilde{\mathcal{J}}_m^{(a,b)}]&=&0,\quad  \{\tilde{\mathcal{J}}_m^{(a,c)},\tilde{\mathcal{J}}_m^{(b,d)}\}=2\delta_{ab}(\delta_{cd}+(1-\delta_{cd})\tilde{\sigma}_3)(\mathcal{H}_m-\mathcal{E}_m),\\
\ [\tilde{\sigma}_3,\mathcal{Q}_m^{(a,c)}]&=&-2i\varepsilon_{ab}\mathcal{Q}_m^{(b,c)},\quad [\tau_3,\tilde{\mathcal{J}}_m^{(a,c)}]=-2i\varepsilon_{ab}\tilde{\mathcal{J}}_m^{(b,c)},\\
{}\{\tilde{\mathcal{J}}_m^{(a,c)},\mathcal{Q}_m^{(b,d)}\}&=&2\delta_{c1}\delta_{d1}\mathcal{X}_{m}^{(a,b)}+2\delta_{c2}\delta_{d2}\varepsilon_{ar}\varepsilon_{bs}\mathcal{X}_{m}^{(r,s)},\\
  {}[\mathcal{X}_{m}^{(a,b)},\mathcal{X}_{m}^{(c,d)}]&=&-2i\left(\varepsilon_{bd}\delta_{ac}\tilde{\sigma}_3+\varepsilon_{ac}\delta_{bd}\tau_3\right)\mathcal{H}_m(\mathcal{H}_m-\mathcal{E}_m),\\
 {}[\mathcal{X}_{m}^{(a,b)},\mathcal{Q}_{m}^{(c,d)}]&=&-2i\left(\varepsilon_{bc}\delta_{d1}\tilde{\mathcal{J}}_{m}^{(a,2)}-\delta_{bc}\delta_{d2}\varepsilon_{ar}\tilde{\mathcal{J}}_{m}^{(r,1)}\right)\mathcal{H}_m,\\
{}[\mathcal{X}_{m}^{(ab)},\tilde{\mathcal{J}}_{m}^{(c,d)}]&=&-2i\left(\varepsilon_{ac}\delta_{d1}\mathcal{Q}_{m}^{(b,2)}-\delta_{ac}\delta_{d2}\varepsilon_{br}\mathcal{Q}_{m}^{(r,1)}\right)\left(\mathcal{H}_m-\mathcal{E}_m\right),\\
{}[\tilde{\sigma}_3,\mathcal{X}_{m}^{(a,b)}]&=&-2i\varepsilon_{bc}\mathcal{X}_{m}^{(a,c)},\quad [\tau_3,\mathcal{X}_{m}^{(a,b)}]=-2i\varepsilon_{ac}\mathcal{X}_{m}^{(c,b)},
\qquad a,b,c,d,r,s = 1,2.\label{shapeall}
\end{eqnarray}
The superalgebra (\ref{PDMWITTEN})-(\ref{shapeall}) is nonlinear again. 
Its linearity is restored in subspaces of fixed energy 
where $\mathcal{H}_m$ acquires constant value. 

The complete set of the fermionic supercharges can be written in the current notation as $\tilde{\mathcal{J}}^{(a,b)}_m$, $i\mathcal{P}\tilde{\mathcal{J}}^{(a,b)}_m$, $\mathcal{Q}^{(a,b)}_m$, and $\mathcal{P}\mathcal{Q}^{(a,b)}_m$. Having in mind that the parity commutes with $\mathcal{Q}^{(a,b)}_m$ while it anticommutes with $\tilde{\mathcal{J}}^{(a,b)}_m$, the structure of the complete $N=16$ superalgebra that would include both local and nonlocal fermionic operators can be found directly from the relations (\ref{PDMWITTEN})-(\ref{shapeall}). In the complete superalgebra, there would also appear bosonic operators $\mathcal{P}\tau_3$ and $\mathcal{P}\tilde{\sigma}_3$.
It is worth noticing that supersymmetric structure of \textit{nonrelativistic} systems with 
two distinct sets of supercharges (both local and nonlocal) was discussed e.g. in 
\cite{AndrianovSokolov} and \cite{trisusy}. $N=8$ extended supersymmetry was considered for three-dimensional Schr\"odinger-Pauli equation in \cite{NikitinNiederle}. In \cite{plyushchayR}, superalgebraic structure based on nonlocal supercharges was considered for one-dimensional nonrelativistic systems.

\subsection{The non-relativistic systems  with position-dependent mass by coordinate transformation}
The same superalgebra (\ref{PDMWITTEN})-(\ref{shapeall}) can be obtained when (\ref{hDg}) and (\ref{J}), corresponding to a generic (diagonal) metric tensor (\ref{metric}), are considered and reduced to the fixed subspaces of $J_3$. We remind that (\ref{hDg}) and (\ref{J}) can be obtained from (\ref{hD}) and (\ref{Jpm}) by the change of coordinates. In this context, let us mention that there appears an interesting class of physical systems where the operators (\ref{hDg}) and (\ref{J}) can appear rather naturally. The Schr\"odinger operators with the 
position-dependent mass $\Sigma(x_1)$,
\begin{equation}\label{pdmh}
 -\frac{1}{\Sigma(x_1)^{1/4}}\partial_{x_1}\frac{1}{\sqrt{\Sigma(x_1)}}\partial_{x_1}\frac{1}{\Sigma(x_1)^{1/4}}+V(x_1),
\end{equation}
emerge in condensed matter physics, e.g. in description of the semiconductors or 
heterostructures \cite{PDMmanybodyapprox}, \cite{PDMheterostructures}. 
These systems have attracted lots of attention in the literature, see 
\cite{PDMSchrodinger}-\cite{Midya} and references therein. 

Let us pick up Dirac 
Hamiltonian (\ref{hDg}). The square of the operator 
in the subspace of fixed angular momentum $J_3=m$ can be written as 
\begin{equation}\label{Hm}
 H_{m}=-\frac{1}{g_{11}^{1/4}}\partial_{x_1}\frac{1}{\sqrt{g_{11}}}\partial_{x_1}\frac{1}{g_{11}^{1/4}}+\left(\frac{m}{\sqrt{g_{22}}}+c_3g_{3}\right)^2-\sigma_3 \frac{1}{\sqrt{g_{11}}}\left(\frac{m}{\sqrt{g_{22}}}+c_3g_3\right)'.
\end{equation}
We can see that the diagonal entries of the matrix operator are of the 
type (\ref{pdmh})  provided that we fix $\Sigma(x_1)\equiv g_{11}(x_1)$.
Let us present an example of a solvable shape-invariant system with position
dependent mass. Since the effective mass in (\ref{pdmh}) is 
required to be bounded and positive for all $x_1$,
we fix the coefficients in (\ref{hDg}) as
\begin{equation}\nonumber
 g_{22}=sn(x_1)^2, \quad g_{11}=dn(x_1)^2,\quad g_3=c_3\frac{cn(x_1)}{sn(x_1)},\quad x_1\in(0,2K).
\end{equation}
It complies with (\ref{g1g}) for $c_1=-1$. Hence, we deal with quantum 
system which possesses $so(3)$ potential algebra. The explicit form of the 
supercharge $Q_{m}^{(1)}$ reads
\begin{equation}\label{Qexample} 
Q_{m}^{(1)}=i\sigma_1\frac{1}{\sqrt{dn(x_1)}}\partial_{x_1}\frac{1}{\sqrt{dn(x_1)}}+\sigma_2\frac{m}{sn(x_1)}+c_3\,\sigma_2\frac{cn(x_1)}{sn(x_1)}.
\end{equation}
The nonrelativistic Hamiltonian $H_m$ acquires the following form
\begin{equation}\label{hpdm}
 H_m=-\frac{1}{dn(x_1)^{\frac{1}{2}}}\partial_{x_1}\frac{1}{dn(x_1)}\partial_{x_1}\frac{1}{dn(x_1)^{\frac{1}{2}}}+\left(\frac{m+c_3\, cn(x_1)}{sn(x_1)}\right)^2+\sigma_3 \frac{c_3+m\, cn(x_1)}{sn(x_1)^2}.
\end{equation}
The effective mass of the nonrelativistic particle is equal to 
$dn^2(x_1)$ and is strictly positive. Notice that it coincides with the mass term  used in \cite{Midya} for $k=1$, $\Sigma(x_1)=dn^2(x_1,1)=\mbox{sech}^2x_1$. The particle lives in the finite 
interval in presence of the external potential. We require the wave 
functions of $H_m$ to be regular and vanishing at the borders of the 
interval. 

The explicit form of the operators $j_{m,\pm}$ reads
\begin{eqnarray}\label{opssphere}
j_{m}^{\pm}&=&i\left(\frac{1}{\sqrt{dn(x_1)}}\partial_{x_1}\frac{1}{\sqrt{dn(x_1)}}\mp\frac{\left( 2m+1\right)cn(x_1)+2c_3+\sigma_3}{2sn(x_1)}\right).
\end{eqnarray}
We will consider $m-1> c_3>0$. For these values of $m$, the potential in 
$H_m$ is confining, i.e. it diverges to infinity at the boundaries.  
Energy levels of $H_m$  are doubly degenerate including the non-zero 
ground state energy $\mathcal{E}_{m}=\left(m+\frac{1}{2}\right)^2-c_3^2$.
Doublet of ground states is formed by $\underline{\psi}_{m}^+$ and 
$\sigma_3\underline{\psi}_{m}^{+}$ where explicitly
\begin{equation}\label{examplehighestvector}
 \underline{\psi}_m^+=\sqrt{dn(x_1)}sn(x_1)^{\frac{1}{2}+m}\left(\frac{1-cn(x_1)}{1+cn(x_1)}\right)^{\frac{c_3}{2}}\left(\begin{array}{c}\sqrt{\frac{1}{2}-c_3+m}\left(\frac{1-cn(x_1)}{1+cn(x_1)}\right)^{\frac{1}{4}}\\-\sqrt{-\frac{1}{2}-c_3-m}\left(\frac{1-cn(x_1)}{1+cn(x_1)}\right)^{-\frac{1}{4}}\end{array}\right).
\end{equation}
The vector complies with the required boundary conditions.
The action of $j_m^-$ on (\ref{examplehighestvector}) preserves regularity 
of wave function provided that $m$ is an integer (semi-integer) and $c_3$ is a 
semi-integer (integer). The $N=2$ supersymmetry of $H_m$ is 
spontaneously broken; the supercharges $Q_{m}^{(1)}$ and 
$i\sigma_3Q_{m}^{(1)}$ do not annihilate ground states, but interchange them mutually. 

The superpartner Hamiltonians $H_m$ and $H_{m+1}$ are spectrally almost 
identical up to the energy level $\mathcal{E}_{m}$ which is missing in the 
spectrum of $H_{m+1}$. Hence, the 
extended operator $\mathcal{H}_m$ in (\ref{PDMextended}) has four-fold 
degeneracy of the energy levels up to the ground state which is doubly 
degenerate. The subspaces with fixed $m$ are invariant with respect to the action of the operators $\mathcal{Q}_m^{(a)}$. The operators $\mathcal{J}_m^{(a)}$ annihilate the 
ground states of $\mathcal{H}_m$, see in Fig. 2 for illustration.

\newsavebox{\pdmspectrum}        
    \savebox{\pdmspectrum}{          
    \scalebox{1}{
    \includegraphics[scale=1]{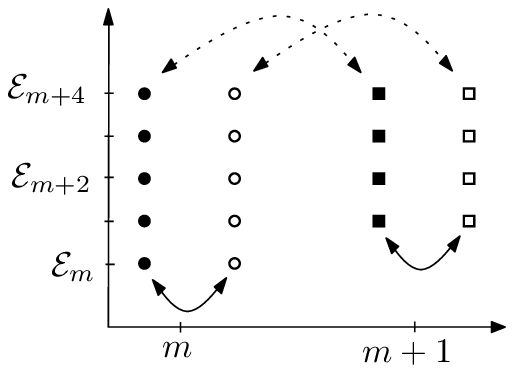}
    }
    } 
\newsavebox{\pdm}        
    \savebox{\pdm}{          
    \scalebox{1}{
    \includegraphics[scale=.6]{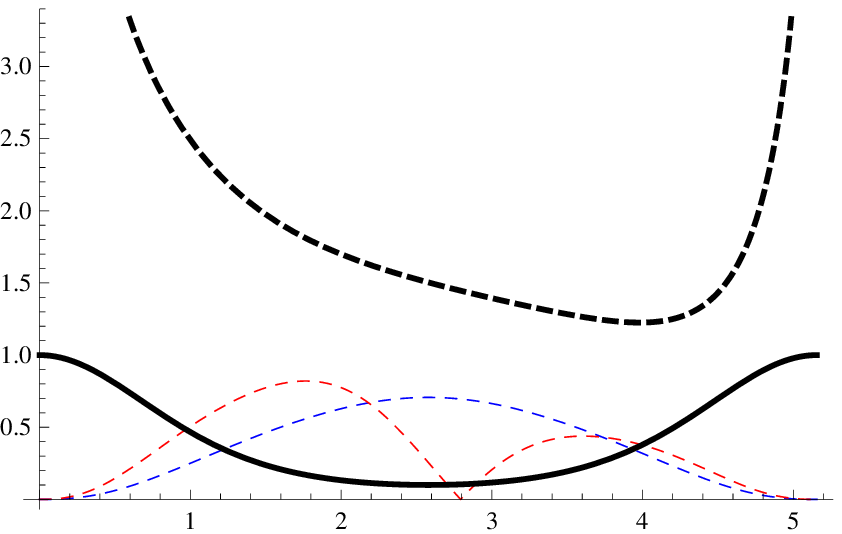}
    }
    }
\begin{figure}[h!]\begin{center}\begin{tabular}{cc}
 \usebox{\pdmspectrum}&\usebox{\pdm}
\end{tabular}\label{pdm}
\caption{(Color on-line) Left: the spectrum of $\mathcal{H}_m$. The operator $\mathcal{Q}_{m}^{(1)}$ 
(solid line) keeps the
doublets of $H_m$ (dots) or $H_{m+1}$ (squares) invariant. 
$\mathcal{J}_{m}^{(1)}$ transforms 
eigenstates of $H_{m}$ into
eigenvectors of ${H}_{m+1}$ and vice versa. Right: probability density of the 
ground state (blue dashed line) and the first excited states (red dashed 
line) of $\frac{\mathbf{1}+\sigma_3}{2}H_{m}$.  Solid black line 
corresponds to $M(x_1)=dn(x_1)^2$, dashed black line corresponds to the 
logarithm of the potential. We fixed $m=2$, $c_3=\frac{1}{2}$ and 
$k=0.9$.}\end{center}\end{figure}

\section{Discussion}
\subsection{Dirac oscillator} 
The quantum settings presented in section \ref{carved} were based on the 
potential algebra $so(3)$. Let us make few comments on the quantum systems 
whose integrals of motion form either the oscillator algebra or $so(2,1)$. 
When $c_1=0$, we have 
$g_2=\frac{1}{x_1}$ and $g_3=c_3x_1$. Substituting these coefficients into (\ref{hD}), we get the Hamiltonian 
\begin{equation}\label{Dosc}
 h_D=i\sigma_1\partial_{x_1}+\sigma_2\frac{J_3}{x_1}+c_3 \sigma_2x_1,\quad x_1>0.
\end{equation}
which coincides with the radial part of the well known Dirac oscillator Hamiltonian introduced by Moshinsky and Szczepaniak \cite{DiracOscillator}.
It has integrals of motion in the following form,
\begin{equation} 
 J_{\pm}=ie^{\pm ix_2}\left(\partial_{x_1}-\frac{\left(\pm J_3+\frac{1}{2}\right)}{x_1}\pm c_3 x_1\pm \frac{\sigma_3}{2x_1}\right).
\end{equation}
The eigenfunctions of (\ref{Dosc}) are required to be vanishing at 
$x_1=0$. The system has bound states as long as $c_3\neq 0$. They can be 
generated from the lowest weight vectors $\psi_{m}^-$, 
\begin{equation} 
 \psi_{m}^-=e^{i m x_2}e^{\frac{1}{2}c_3x_1^2}x_1^{-m}\left(-i\sqrt{c_3\left(-\frac{1}{2}+m\right)},\sqrt{2}c_3x_1\right)^T,
\end{equation} 
and comply with the required boundary condition provided that $c_3<0$ and 
$m$ is a semi-integer lower or equal to $\frac{1}{2}$. 

In general, the solution of the stationary equation with (\ref{Dosc}) are
not determined uniquely by the requirement of square 
integrability. The singularity of the potential
at $x_1=0$ causes that the square integrability is 
not sufficient to determine the wave functions. 
Our requirement of regularity of the wave functions at the origin removes the
ambiguity by fixing one of the self-adjoint extensions of $h_D$.   
In \cite{Znojil}, different way of dealing with singular potentials was discussed. 
The potential of Calogero Hamiltonian was regularized by the 
complex shift of the coordinate. Its hermicity was violated, however, the complex shift of 
coordinate allowed for additional set of solutions and spectrum remained real (see also 
\cite{MikhailPT} for a recent reference where the technique was used).
This approach should be applicable in the context of Dirac Hamiltonian (\ref{Dosc}) 
as well as in other quantum systems described by (\ref{hD}) with singular coefficients 
$g_{2}$ and $g_3$. 

\subsection{$\mathbf{so(2,1)}$ and BTZ black hole -like system}
Finally, let us discuss the case where the quantum system possesses 
integrals of motion that form $so(2,1)$. We shall consider $x_2$ as the 
time-like coordinate. In order to do so, we perform the substitution 
$x_2\rightarrow -it$. It makes the replacement $dx_2^2\rightarrow -dt^2$ 
in the line element $ds^2=g_{\mu\nu}dx^{\mu}dx^{\nu}$ and, hence, the sign 
of the lower component of $g_{\mu\nu}$ gets changed accordingly. This 
substitution transforms Dirac Hamiltonian (\ref{hDepsilon1}) into 
(\ref{hDepsilon-1}) (the flat space metric is now $\eta_{\mu\nu}=diag(1,-1)$) 
while keeping the condition (\ref{g1g}) unchanged. As an example, let us 
consider the space with the metric tensor in the 
following form 
\begin{equation}\label{almostBTZg} 
g_{\mu\nu}=\mbox{diag}\left(N^{-2},-N^2\right),\quad N=N(x_1),
\end{equation} 
where we leave the function $N$ unspecified at the moment. Let us comment 
that metric tensors of similar type appear in description of 
black holes \cite{2D black hole}.
The Dirac Hamiltonian (\ref{hDepsilon-1}) associated with 
(\ref{almostBTZg})  reads
\begin{equation}\label{almostBTZ}
 h_D=iN\sigma_1\partial_{x_1}+\sigma_2\frac{\partial_{t}}{N}+i\sigma_1\frac{N'}{2}+c_3\sigma_2 N'.
\end{equation}
It has integrals of motion (\ref{J}), 
\begin{equation}
 J_{\pm}=ie^{\pm t}\left(N\partial_{x_1}\mp N'\partial_t\mp \frac{c_3}{N}\mp\frac{\sigma_3}{2N}\right).
\end{equation}
They, together with $J_3=\partial_t$, form the $so(2,1)$ algebra provided 
that the components of the metric tensor (\ref{almostBTZg}) solve 
(\ref{g1g}). We find very interesting that there is a solution of (\ref{g1g}) where the function $N$ acquires the following form 
\begin{equation}
N^2=-1+x_1^2.
\end{equation}
The metric (\ref{almostBTZg}) then coincides with the one of BTZ black 
hole \cite{BTZ} as long as the black hole has unit mass and zero angular 
momentum. The particle lives on the straight line with fixed angular coordinate
which terminates in the singularity. This coincidence 
suggests that appropriate (higher-dimensional) modification of our current 
algebraic approach could be particularly helpful in the analysis of Dirac 
fermions in the gravitational background of BTZ black hole.  

There appeared remarkable results recently where the geometrical 
properties of the settings with horizon were encoded into the spatial part of the 
metric tensor \cite{Iorio}. This observation made it possible to consider black hole simulations 
in the strained graphene. Analysis of these 
systems in the context of the potential algebras could be very 
interesting.

\subsection{Outlook}
In the current work, we considered potential algebras of the $(2+1)$-
dimensional Dirac-like operator (\ref{hD}). The explicit form the operator was found
that possessed Lie algebra $so(3)$, $so(2,1)$ or 
oscillator algebra of integrals of motion. 
It should be mentioned that we do not make any statements about generality of the found representations; the work was rather focused on the applications of the potential algebras.
The results can be understood in the context of classification
of realizations of the low-dimensional Lie algebras. Here, let us mention \cite{popovych} where representations in terms of differential operators without matrix degree of freedom were analyzed.
The analysis of the generic form of the ladder operators $J_{\pm}$ that would satisfy the algebra (\ref{J3})-(\ref{algebraansatz}) would be desirable.
It would be a step towards  the extension of the general results obtained in \cite{popovych} to the operators with matrix coefficients.

The structure of the potential algebra, its rank in particular, was fixed  
by the ansatz (\ref{algebraansatz}) and by the explicit form of the 
Dirac Hamiltonian (\ref{hD}). It would be possible to extend the current approach 
to higher-dimensional settings by considering potential algebras of higher 
rank. Such algebraic structures could be used effectively in the analysis 
of quantum systems that live in spaces with non-trivial geometry. 
It might be applied 
in construction of the new, shape-invariant, two-dimensional Dirac 
Hamiltonians.  As it 
was suggested above, it could be also useful in the analysis of Dirac particle 
in the gravitational background of BTZ black hole. 

In the considered framework, the operators $J_{\pm}$ were required to 
commute with the Hamiltonian, i.e. they  reflected degeneracy of its 
spectrum. The concept of dynamical algebras (or spectrum generating 
algebras) represents a natural extension of the current approach. Relaxing 
the condition $[J_{\pm},h_D]=0$, we could consider a broader class of 
quantum systems. 
A possible hint on how to modify the commutator is provided by the 
shape-invariance in the nonrelativistic quantum mechanics, where $H^-_{m+1}q_{m}-q_{m}H^-_{m}=r(m)q_{m}$ with $r(m)$ being a constant. Having in mind the reduction 
(\ref{hrestricted}) and (\ref{j}), the commutator is suggested to be   $[h_D,J_{\pm}]\sim J_{\pm}. $
The ansatz for 
the operators $J_{\pm}$ should be extended appropriately to the matrix 
operator with generally non-vanishing anti-diagonal components.

It is worth mentioning in this context that a different approach to  
dynamical symmetries of nonrelativistic systems was introduced in 
\cite{SGANegro}. Dynamical algebra of Poschl-Teller system was treated as 
a potential algebra of a suitably modified Hamiltonian. Implementation of 
this approach to Dirac operators could be fruitful as well.

\appendix
\section*{Appendix A}
We shall show that for the representations of $so(3)$ generated by 
(\ref{Jpm}) from the lowest weight vectors, the coupling constant $c_3$ 
has to be integer (half-integer) as long as $m$ is  half-integer 
(integer). Only in this case, the representations can be finite-
dimensional. 

We set $g_2=\sin^{-1} x_1$ and $g_3=c_3\cot x_1$ in (\ref{Jpm}).
The lowest weight vector, the kernel of $J_-$, acquires the form
 $$\psi_{m}^{-}=e^{imx_2}\sin^{\frac{1}{2}+m}x_1\left(\beta_1\tan^{\frac{1}{2}+c_3}\frac{x_1}{2},\beta_2\tan^{-\frac{1}{2}+c_3}\frac{x_ 1}{2}\right)^T,$$
where $T$ denotes transposition.
The actual value of the constants $\beta_1$ and $\beta_2$ 
is not important at the moment. Next we define a diagonal matrix $G$,
\begin{equation}\label{G}\nonumber
 G=\sin^{\frac{1}{2}}x_1\mbox{diag}\left(\tan^{\frac{1}{2}+c_3}\frac{x_1}{2},\tan^{-\frac{1}{2}+c_3}\frac{x_1}{2}\right)
\end{equation}
We use this matrix to transform the ladder operator $J_+$ and the lowest-weight vector to the form that will be convenient for the forthcoming calculation, 
\begin{eqnarray}\label{gaugedJ2}
 G^{-1}J_+G&=&ie^{ix_2}g_2\sqrt{-1+g_2^2}\left(\frac{1}{g_2'}\partial_{x_1}+\frac{J_3}{g_2}\right),\quad
\psi_{m}^-=e^{imx_2}G\, \left(\begin{array}{r}\alpha^- g_2^{m-1+2c_3}w^{2c_3-1}\\\beta^-g_2^{m+1+2c_3}w^{2c_3+1} \end{array}\right), 
\end{eqnarray}
where $w=1+\sqrt{1-\frac{1}{g_2^2}}$. Let us compute how does $J_+$ act on the generic wave function $\xi=e^{in 
x_2}G\,\left(^{g_2^{r}w^s}_0\right)$. Using (\ref{gaugedJ2}), we 
can calculate immediately
\begin{equation}\label{actionJ}
J_+e^{in x_2}G\,\left(\begin{array}{c}g_2^{r}w^s\\0\end{array}\right)=ie^{i(n+1)x_2}G\,\left(\begin{array}{c}-(n+r-s)w^{s-1}g_2^{r-1}+g_2^{r+1}w^s(n+r)\\0\end{array}\right).
\end{equation}
This formula helps to understand qualitatively the structure of the 
$(J_+)^k\xi$ (with $k$ being a positive integer); we can find coefficients 
at the highest and the lowest power of $g_2$. The coefficient of the term 
$g_2^{r-k}w^{s-k}$ is $(-1)^{k}\prod_{l=0}^{k-1}(n+l-1+r-s)$ while the 
coefficient of the term $g_2^{r+k}w^s$ is $\prod_{l=0}^{k-1}(n+r+2l)$. 

Now, let us identify the nonzero element of $\xi$ with the upper component 
of $\psi_m^-$ in (\ref{gaugedJ2}),
i.e. we fix $n=m$, $r=m-1+2c_3$ and $s=2c_3-1$. The requirement that 
$\psi_{m}^-$ is annihilated by a specific power of $J_+$ implies that the 
above derived coefficients have to vanish for a specific value of $k$. 
This happens provided that 
\begin{equation}\label{conditions}\nonumber
2m-1\in \mathbb{Z}\leq0,\quad -(m+c_3)+\frac{1}{2}\in \mathbb{Z}\geq0.
\end{equation}  
The first relation restricts $m$ to be integer or half-integer while the 
second relation tells that $c_3$ has to be half-integer or integer, 
respectively.

\section*{Appendix B}
Let us review briefly how the Dirac Hamiltonian for the mass-less particle in 
the curved space can be constructed. We will consider two distinct scenarios that are distinguished by the sign $\varepsilon$ in the metric tensor $g_{\mu\nu}$,
\begin{equation}\nonumber
g_{\mu\nu}=\left(\begin{array}{cc} g_{11}(x_1)&0\\0&\varepsilon\, g_{22}(x_1)\end{array}\right).
\end{equation}
We suppose that $g_{11}(x_1)\geq0$ and  $g_{22}(x_1)\geq0$.
For $\varepsilon=-1$, we deal with $(1+1)$ dimensional space-time, considering the coordinate $x_2$ to be time-like. When $\varepsilon=1$, both coordinates are space-like and the metric describes curved space. In both cases, Dirac equation can be written in the following form \cite{Birrel}, 
\begin{equation}\label{birreleq}
 i\gamma^{\mu}\left(\partial_{\mu}+\Omega_{\mu}\right)\psi=\lambda\psi,\quad \mu=1,2,
\end{equation} 
where $\lambda$ corresponds to the mass of the particle for $\epsilon=-1$ whereas it represents eigenvalues of Dirac Hamiltonian for $\epsilon=1$. In the latter case, (\ref{birreleq}) can be identified with the stationary equation. $\Omega_{\mu}$ is spin connection and $\gamma^{\mu}$ are generalized gamma matrices which satisfy $\{\gamma^{\mu},\gamma^{\nu}\}=2g^{\mu\nu}$. These quantities can be computed with the use of tetrad formalism. 
The tetrads $e_{x_{\mu}}^a$ are related to the metric tensor in the following manner,
$$e_{x_{\mu}}^ae_{x_{\nu}}^b\eta_{ab}=g_{\mu\nu}. $$
The quantities $e_{x_{\mu}}^a$ and $e^{a}_{x_{\nu}}$ relate locally flat coordinates $y_a$ with the curvilinear 
coordinates $x_\mu$ at a 
given point $X$, $e^{a}_{x_{\mu}}=\left(\partial_{x_{\mu}}{y_a}\right)(X)$ and 
$e_{a}^{x_{\mu}}=\left(\partial_{y_a}{x_{\mu}}\right)(X)$. There holds 
$e_{a}^{x_{\mu}}e^{b}_{x_{\mu}}=\delta_{a}^{b}$. Flat-space metric is identified as $\eta_{ab}=\mbox{diag}(1,\varepsilon)$.
The matrices $\gamma^{\mu}$ are defined as 
\begin{equation}
\gamma^{\mu}=e^{\mu}_a\gamma^a\quad \mbox{where}\quad \{\gamma^{a},\gamma^{b}\}=2\eta^{ab}.
\end{equation} The spin connection $\Omega_{\mu}$ is defined as 
\begin{equation}\nonumber
 \Omega_{\mu}=\frac{1}{4}\omega^{ab}_{\mu}[\gamma_{a},\gamma_b],
\end{equation}
where
\begin{equation}
 \omega^{ab}_{\mu}=e^{a}_{x_{\lambda}}g^{\lambda\tau}\left(\partial_{\mu}e^{b}_{x_{\tau}}-\Gamma_{\mu\tau}^{\kappa}e^{b}_{x_{\kappa}}\right),\quad \Gamma^{\lambda}_{\mu\nu}=\frac{1}{2}g^{\lambda\sigma}\left(\partial_{\nu}g_{\sigma\mu}+\partial_{\mu}g_{\sigma\nu}-\partial_{\sigma}g_{\mu\nu}\right).
\end{equation}

In our present case, the tetrads can be fixed as 
\begin{equation}\nonumber
 e_{x_{\mu}}^a=\left(\begin{array}{cc}\sqrt{g_{11}}&0\\0&\sqrt{g_{22}}\end{array}\right),\quad e_{a}^{x_{\mu}}=\left(\begin{array}{cc}\frac{1}{\sqrt{g_{11}}}&0\\0&\frac{1}{\sqrt{g_{22}}}\end{array}\right).
\end{equation}
Here, the upper index denotes row and the lower index the column of the 
matrix. Non-zero elements of the affine connection are 
$ \Gamma^{1}_{11}=\frac{g_{11}'}{2g_{11}},\quad 
\Gamma^{1}_{22}=-\varepsilon\frac{g_{22}'}{2g_{11}},\quad 
\Gamma^{2}_{12}=\Gamma^{2}_{21}=\frac{g_{22}'}{2g_{22}}
$ and
\begin{equation}\nonumber
 \omega^{12}_{2}=-\omega^{21}_{2}=-\frac{g_{22}'}{2\sqrt{g_{11}g_{22}}}.
\end{equation}
The matrices $\gamma_a$ are defined as $\gamma_1=\sigma_1$,  
$\gamma_2=i^{\frac{1-\varepsilon}{2}}\sigma_2$ and satisfy 
$\{\gamma_a,\gamma_b\}=2\eta_{ab}$. The generalized gamma matrices are 
then
\begin{equation}\nonumber
 \gamma^{x_1}=\sqrt{g_{11}}^{-1}\sigma_1,\quad  
\gamma^{x_2}=\varepsilon \, i^{\frac{1-\varepsilon}{2}}\sqrt{g_{22}}^{-1}\sigma_2.
\end{equation}
Finally, the spin connection is 
\begin{equation}\nonumber
 \Omega_{x_1}=0,\quad \Omega_{x_2}=-\frac{i\left(i^{\frac{1-\varepsilon}{2}}\right)g_{22}'}{4\sqrt{g_{11}g_{22}}}\sigma_3.
\end{equation}
Dirac Hamiltonian for $\varepsilon=1$ acquires the form
\begin{equation}\label{hDepsilon1}
 h^{\varepsilon=1}_D=i\sigma_1\left(\frac{1}{\sqrt{g_{11}}}\partial_{x_1}+\frac{g_{22}'}{4g_{22}\sqrt{g_{11}}}\right)+\sigma_2\frac{i\partial_{x_2}}{\sqrt{g_{22}}},
\end{equation}
while it has the following explicit form for $\varepsilon=-1$
\begin{equation}\label{hDepsilon-1}
 h_D^{\varepsilon=-1}=i\sigma_1\left(\frac{1}{\sqrt{g_{11}}}\partial_{x_1}+\frac{g_{22}'}{4g_{22}\sqrt{g_{11}}}\right)-\sigma_2\frac{\partial_{x_2}}{\sqrt{g_{22}}}.
\end{equation}
We transform the operator $h^{\varepsilon=1}_D$ to the form which is 
manifestly hermitian with respect to the standard scalar product in the 
following manner 
\begin{equation}\label{hDepsilon11}\nonumber
 h_D=(\det g_{\mu\nu})^{\frac{1}{4}}h^{\varepsilon=1}_D(\det g_{\mu\nu})^{-\frac{1}{4}}=\frac{i\sigma_1}{\sqrt{g_{11}}}\partial_{x_1}+\frac{i\sigma_2}{\sqrt{g_{22}}}\partial_{x_2}-\frac{i\sigma_1 g_{11}'}{4 g_{11}^{3/2}}.
\end{equation}
$$ $$
\noindent
{\bf Acknowledgements:} The author thanks F. Correa for fruitful 
discussion on BTZ black holes. He also thanks the referee for constructive comments. The work  was supported by the GA\v CR 
Grant P203/11/P038, Czech Republic.

\end{document}